# On Energy Efficiency and Performance Evaluation of SBC based Clusters: A Hadoop case study

Basit Qureshi[1], Anis Koubaa[1,2]
[1] Robotics and IoT Lab, Prince Sultan University, Saudi Arabia
[2] CISTER/INESC-TEC, ISEP, Polytechnic Institute of Porto, Porto, Portugal
{qureshi,akoubaa}@psu.edu.sa

*Abstract*: Energy efficiency in a data center is a challenge and has garnered researchers interest. In this paper we address the energy efficiency issue of a small scale data center by utilizing Single Board Computer (SBC) based clusters. A compact design layout is presented to build two clusters using 20 nodes each. Extensive testing was carried out to analyze the performance of these clusters using popular performance benchmarks for task execution time, memory/storage utilization, network throughput and energy consumption. Further, we investigate the cost of operating SBC based clusters by correlating energy utilization for the execution time of various benchmarks using workloads of different sizes. Results show that, although the low-cost benefit of a cluster built with ARM-based SBCs is desirable, these clusters yield low comparable performance and energy efficiency due to limited onboard capabilities. It is possible to tweak Hadoop configuration parameters for an ARM-based SBC cluster to efficiently utilize resources. We present, a discussion on the effectiveness of the SBC-based clusters as a testbed for inexpensive and green cloud computing research.

## 1. Introduction

Energy consumption in data centers is a major concern for green computing research. In the 2017 Green peace [1] clean report, the global energy consumption for data centers is estimated to be over 31 GigaWatts, only 2 GigaWatts can be attributed to energy from renewable resources. The NRDA [2] estimated, in the US alone, the data centers consumed 91 billion kilowatts hours (kWh) of energy, which is estimated to increase by 141 billion kWh every year until 2020, costing businesses $13 billion annually in electricity bills and emitting nearly 100 million metric tons of carbon pollution per year. Resource over-provisioning and energy non-proportional behavior of today's servers [3] are two of the most important reasons for high energy consumption of data centers. At the same time, environmental concerns faced by many large-scale cloud computing infrastructure operators, have prompted the need for more energy efficient operation of infrastructure. Recent considerations in energy efficiency [4–10] has improved the understanding and need for more energy efficient Cloud computing technologies.

In order to build a cloud computing cluster with low energy consumption requirements resulting in near-zero carbon footprint, researchers have investigated the use of Single Board Computers (SBC) [11,12]. A SBC is a complete computer built on a single circuit board incorporates a microprocessor(s), memory, I/O as well as multitude of other features required by a functional computer. Typically an SBC is ideally priced at (35 to 80 US$), with power requirements set to be as low as 2.5 Watts and designed in small form factors comparable to a credit card or pocket size. These computers are portable and are capable of running a wide range of platforms including Linux distributions, Unix, Microsoft Windows, Android etc. Researchers have built clusters for high-performance computing research using SBCs [13,14]. A cluster of single board computers has very limited resources and cannot compete with the performance of high end servers. But despite these drawbacks, useful application scenarios exist, where clusters of single board computers are a promising option[2]. This applies in particular to small and medium-sized enterprises as well as for academic purposes like student projects or research projects with limited financial resources.

The Beowulf cluster created at Boise State University [15] was perhaps the earliest attempt at creating a cluster consisting of multiple nodes of SBCs. This cluster is composed of 32 Raspberry Pi Model B computers and offers an alternative in case if the main cluster is unavailable. The Bolzano Raspberry Pi cloud cluster experiment implemented a 300 node Pi cluster [16]. The main goal of this project was to study the process and challenges of building a Pi cluster on such a large scale. The Iridis-Pi project implemented a 64 node Raspberry Pi cluster [17]. Tso et al. [18] built a small-scale data center consisting of 56 RPi Model B boards. The Glasgow Raspberry Pi Cloud offers a cloud computing testbed including virtualization management tools. In 2016, C. Baun in [19] presented the design of a cluster geared towards academic research and student scientific projects building an 8-node Raspberry Pi Model 2B cluster. All of these works demonstrate constructing a cluster using SBCs at an affordable cost to researchers and students. More recently, authors in [20–25] have used SBC devices or clusters in Edge computing scenarios. However, none of these works provide a detailed performance and power efficiency of executing hadoop operations in such clusters.

In this paper, we present a detailed study on design and deployment of two SBC based clusters using the popular and widely available Raspberry Pi and HardKernel Odroid Model Xu-4. The objectives of this study are in three fold: i) To provide a detailed analysis of the performance of Raspberry Pi and Odroid Xu-4 SBCs in terms of power consumption, processing/execution time for various tasks, storage read/write as well as network throughput; ii) To study the viability and cost effectiveness of the deployment of SBC based Hadoop clusters against virtual machine based Hadoop clusters deployed on personal computers; iii) To contrast the power consumption and performance aspects of SBC based Hadoop clusters for Big-Data Applications in academic research. To this end, two clusters were constructed and deployed for an extensive study of the performance aspects of individual SBCs and a cluster of SBCs. Hadoop was deployed on these clusters to study the performance aspects using benchmarks for power consumption, task execution time, I/O read/write latencies and network throughput. In addition to the above, we provide analysis of energy consumption in the clusters, the energy efficiency and cost of operation. The contribution of this paper is as follows

- Design for two clusters using SBCs are presented in addition to a PC based cluster running in the Virtual environment. Performance evaluation of task execution time, storage utilization, network throughput as well as power consumption are detailed.
- Popular Hadoop benchmark programs such as Pi Computation, Wordcount, TestDFSIO, TeraGen, and TeraSort are executed on these clusters and results are compared against a Virtual Machine based cluster using workloads of various sizes.

- An in-depth analysis of energy consumption was carried out for these clusters. The cost of operation is analyzed for all clusters by correlating the performance of task execution times and energy consumption for various workloads.

The remainder of this paper is organized as follows. Section 2 presents related works with details on the ARM-based computing platforms used in this study as well as a review of recent applications of SBCs in High-performance computing and Hadoop based environments. Section 3 presents the design and architecture of the RPi and Xu20 clusters used in this study. Section 4 deals with a comprehensive performance evaluation study of these clusters based on popular benchmarks. Section 5 provides details on the deployment of Hadoop environment on these clusters with a detailed presentation of performance aspects of Hadoop benchmarks for the clusters. Section 6 provides a detailed analysis on the impact of power consumption and CPU temperature on the three clusters. Section 7 provides summary and discussion followed by conclusions in section 8.

## 2. Background

This section is subdivided into two sections. Section A details the SBC platforms used in this study, whereas section B presents related work on SBC based clusters.

### 2.1. The SBC Platforms

Advanced RISC Machine (ARM) is a family of Reduced Instruction Set Computing (RISC) architectures for computer processors that are commonly used nowadays in tablets, phones, game consoles etc. The ARM is the most widely used instruction set architecture in terms of quantity produced [12].

The Raspberry Pi Foundation [26], developed a credit card-sized SBC called Raspberry Pi (RPi). This development was aimed at creating a platform for teaching computer science and relevant technologies at the school level. Raspberry Pi 2B version was released in February 2015 improving the previous development platform by increased processor speed, larger onboard memory size as well as newly added features. Although the market price, as well as the cost of energy consumption of an RPi, are low, the computer itself has many limitations in terms of shared compute and memory resources. In summary, the RPi is a very affordable platform with low cost and low energy consumption [27,28]. The major drawback is the compute performance. Recent experiments in distributed computing have shown that this can be rectified by building a cluster of many RPi computers.

The Hardkernel Odroid platform. ODROID-XU-4 [29] is a newer generation of single board computers offered by HardKernel. Offering open source support, the board can run various flavors of Linux, including Ubuntu, Ubuntu MATE and Android. XU-4 uses Samsung Exynos5 Quad-core ARM CortexTM-A15 Quad 2Ghz and CortexTM-A7 Quad 1.3GHz CPUs with 2Gbyte LPDDR3 RAM at 933MHz. The Mali-T628 MP6 GPU supports OpenGL 3.0 with 1080p resolution via standard HDMI connector. Two USB 3.0 ports, as well as a USB 2.0 port, allows faster communication with attached devices. The power-hungry processor demands 4.0 amps power supply with power consumption of 2.5 Watts (idle) and 4.5 Watts (under load). Odroid XU-4 priced at $79, is slightly expensive compared to Raspberry Pi 3B, nevertheless the improved processing power although demanding more power provides tradeoff with improved performance, task execution time as well as better I/O read and write operations. Table 1 shows a summary comparison of Raspberry Pi 2B and Ordoid Xu-4 SBCs.

**Table 1.** Features of Raspberry Pi Model 2B and HardKernel Odroid Xu-4

|  | RPi Model 2B | Odroid XU-4 |
|---|---|---|
| Processor (CPU) | 0.9 GHz quad core ARM Cortex-A7 | Samsung Exynos5 Octa ARM Cortex-A15 (@ 2.0 GHz) and Cortex-A7 (@1.3GHz) CPUs |
| GPU | Broadcom Video Core IV Multimedia Graphics co-processor | Mali T628 Open GL 3.0 |
| Onboard RAM | 256 KB L2 cache 1 GB SDRAM at 400MHz | 2GB LPDDR3 at 933MHz |
| Ethernet / Network | 10/100 MB Ethernet RJ45 Jack | 10/100/1000 MB Ethernet RJ45 Jack |
| Storage | Micro SD Card | Micro SD Card and eMMC 5.0 flash storage |
| Audio/Video | 3.5 mm jack and HDMI | HDMI (standard) supports 1080p video |
| Power Consumption | 3.2 W (idle) 3.8 W (under load) | 2.5 W (idle) 4.5 W (under load) |
| USB Ports | 4 USB 2.0 | 1x USB 2.0, 2x USB 3.0 |
| Released | February 2015 | 2015 |
| Price (US$) | 35$ | 79$ |

### 2.2. The SBC Cluster Projects

The Beowulf cluster created at Boise State University in 2013 [15] created a cluster consisting of multiple nodes of SBCs. It was built for collaboratively processing sensor data in a wireless sensor network. This cluster is composed of 32 Raspberry Pi Model B computers and offers an alternative in case if the main cluster is unavailable. This work documents the cluster construction process and provides information on the clusters performance and power consumption. The researchers present the compute performance of single RPi and an Intel Xeon III based server using the Message Passing Interface libraries (MPI) running computation of the value of pi using Monte Carlo method. They first compare a single RPi against 32 RPi's organized in a cluster and report improvement of the speed up as well as a decrease in the execution time. However, when the RPi Cluster is compared to the Intel Xeon server, the Xeon server performs 30 times better in terms of execution time.

The Bolzano Raspberry Pi cloud cluster experiment implemented a 300 node Pi cluster [16]. The main goal of this project was to study the process and challenges of building a Pi cluster on such a large scale. The researchers demonstrate how to setup and configure the hardware, the system, and the software. In their work, Abrahamsson et. al. presented applications of this cluster as a testbed for research in an environmentally friendly, green computing. Furthermore, they also considered using this cluster to be deployed as a mobile data center. Although the focus of this work is on the design and deployment of the cluster using Raspberry Pi Computers, the work lacks detailed performance analysis of the cluster using popular performance benchmarks, as presented in this work.

The Iridis-Pi project implemented a 64 node Raspberry Pi cluster [17]. Commonly known as the Lego super-computer, the work presents design and deployment of the raspberry pi cluster using Lego blocks in a compact layout. They present a detailed analysis of performance in terms of execution time, network throughput, as well as I/O, read/write. The cluster computes performance is measured using the HPL Linpack benchmark which is popularly used to rank the performance of supercomputers. The network performance was measured using a Message Passing Interface (MPI) to communicate between the Raspberry Pi. Researchers argue that although the cluster cannot be used in conventional supercomputing environments due to its lacking performance, however the low cost, energy efficient, open source architecture, allows future academics and researchers to consider the use of such clusters.

Tso et al. [18] built a small scale data center consisting of 56 RPi Model B boards. The Glasgow Raspberry Pi Cloud offers a cloud computing testbed including virtualization management tools. The primary purpose of this research was to build a low-cost testbed for cloud computing resource management and virtualization research areas to overcome the limitations of simulation-based studies. The work compares the acquisition cost, electricity costs and cooling requirements of the cluster of single board computers with a testbed of 56 commodity hardware servers. Although the work presented provides a testbed for cloud computing research, no further details are available on the performance comparison of this work. In this paper, we present a detailed analysis of SBC based cluster's performance attributes in Hadoop environment.

Cubieboards [30] single board computer presented a Hadoop cluster of Eight nodes. They compared the performance of Raspberry Pi Model A and Model B against the Cubieboard and concluded the Cubieboard is better suited for Hadoop deployment due to the faster CPU at 1 GHz as well as a bigger main memory of 1 GB. The authors provide a complete step-by-step guide for deploying Hadoop on the cubie board platform for students and enthusiasts. They demonstrated the use of Wordcount program on a large 34 Gigabyte file obtained from Wikipedia. Although the demonstration shows deployment of the Hadoop cluster, the authors do not present any performance analysis results. Kaewkas and Srisuruk [31] at Suranaree University of Technology built a cluster of 22 Cubieboards running Hadoop and Apache Spark. They performed various tests studying the I/O performance and the power consumption of the cluster. They conclude that a 22-node cubie-board based cluster is enough to perform basic Big-Data operations within an acceptable time. In 2016, C. Baun in [19] presented the design of a cluster geared towards academic research and student science projects. They argue for the case of the physical representation of the cloud infrastructure to the students which may not be accessible in a public cloud domain. They built an 8-node Raspberry Pi Model 2B cluster and study the performance aspects including computation time, I/O reads and writes as well as Network throughput.

In 2018, [12] deployed clusters using Raspberry Pi and other SBCs as a edge computing device. The proposed clusters would be used in the context of smart city applications. In 2018, researchers [21] introduced a Lightweight Edge Gateway for the Internet of Things (LEGIoT) architecture. It leverages the container based virtualization using SBC devices to support various Internet of Things (IoT) application protocols. They deploy the proposed architecture on a SBC device to demonstrate functioning of an IoT gateway. Morabito [20], in 2017, provided a performance evaluation study of popular SBC platforms. They compare various SBCs using Docker container virtualization in terms of CPU, Memory, Disk I/O, Network performance criterion. Also recently in 2018, authors in [23] conducted an extensive performance evaluation on various embedded microprocessors systems including Raspberry Pi. They deploy Docker based containers on the systems and analyze the performance of CPU, Memory and Network communication on the devices.

The low-cost aspect of an SBC makes it attractive for students are well as researchers in academic environments. It remains to be seen how the SBCs perform when deployed in Hadoop clusters. Further investigation is needed to understand the cost of energy and efficiency of executing Hadoop jobs in these clusters.

In this paper, we address this gap in literature by:
- Providing a detailed analysis of the performance of Raspberry Pi and Odroid Xu-4 SBCs in terms of power consumption, processing/execution time for various tasks, storage read/write as well as network throughput.
- Studying the viability and cost-effectiveness of the deployment of SBC based Hadoop clusters against virtual machine based Hadoop clusters deployed on personal computers.
- Analyzing the power consumption and energy efficiency of SBC based Hadoop clusters for Big-Data Applications.

To this end we deploy three clusters to extensively study the performance of individual SBCs as well as the Hadoop deployment, using popular performance benchmarks. We compare power consumption, task execution time, I/O read/write latencies as well as network throughput. Furthermore, we provide a detailed analysis and discussion on the energy efficiency and cost of operating these clusters for various workloads. The next section presents details about the construction of the clusters.

## 3. Design and Architecture of the SBC Clusters

The first cluster, called RPi-Cluster, is composed of 20 Raspberry Pi Model 2B Computers connected to a network. The second cluster, called Xu-20, is composed of 20 Odroid XU-4 devices in the same network topology. The third cluster named HDM, is composed of four regular PCs running Ubuntu in the Virtual environment using VMware Workstation[32]. To maintain similarity in network configuration, all the clusters follow the same star topology with a 24-port Giga-bits-per-second smart managed switch acting as the core of the network as can be seen in Figure 1. Each node (RPi, Xu-4 or PC) connects a 16-port Ethernet switch that connects to the core switch. Currently, five nodes connect to each switch allowing further scalability of the cluster. The master node, as well as the uplink connection to the Internet through a router, is connected to the core switch. The current design allows easy scalability with up to 60 nodes connected in the Cluster that can be extended up to 300 nodes. Table 2 presents a summary of the cluster characteristics.

### 3.1. Components and the Design of the DM-Clusters

Each cluster is composed of a set of components including SBCs, Power supplies, network cables, Storage modules, connectors, and cases. Each SBC is carefully mounted with Storage components. All the Raspberry Pi computers are equipped with 16 GB Class-10 SD cards for primary bootable storage. The Odroid Xu-4 devices are equipped with 32GB eMMCv5.0 modules. All the SBCs are housed in a compact layout racks using M2/M3 spacers, nuts, and screws. The racks are designed to house 5 SBCs per rack for easy access and management.

Currently, each Raspberry Pi computer is individually supplied by the 2.5Amp power supply; each Odroid Xu-4 computer is supplied by a 4.0Amp power supply that provides ample power for running each node. All the power supplies are connected to the Wattsup Pro .net power supply meter for measuring power consumption.

Each SBC's network interface is connected to a Cat6e Ethernet cable through the RJ-45 Ethernet connector. All Ethernet cables connect to the 16-port Cisco switches which connect to a Gigabit Core switch. An Internet router, as well as the Master PC running Hadoop namenode, is connected to the network. The HDM Cluster is composed of four PCs all connected in the same network topology as of the other clusters. Each PC is equipped with an Intel i7 4th Gen Processor with 3.0Ghz Clock speed, 8 GB RAM and 120 GB Solid State Disk Drive for storage. Each of these PC's is equipped with a 400W power supply and connects to the Ethernet Switch. The purchase cost of all components of the RPi, Xu20 and HDM Clusters was $1,300, $2,700 and $4,200 respectively. The Network and Power reading equipment cost is approximately $450.

### 3.2. Raspbian and Ubuntu MATE Image installation

For the RPi Cluster, we built the RPi Image. The Raspbian OS image is based on Debian that is specifically designed for ARM processors. Using Raspbian OS for RPi is easy with minimal configuration settings requirements. Each individual RPi is equipped with a SanDisk Class 10, 16 GB SD card capable of up to 45MB/s read as well as up to 10MB/s write speeds available at a cost of US$15. We created our own image of the OS which was copied on the SD cards. Additionally, Hadoop

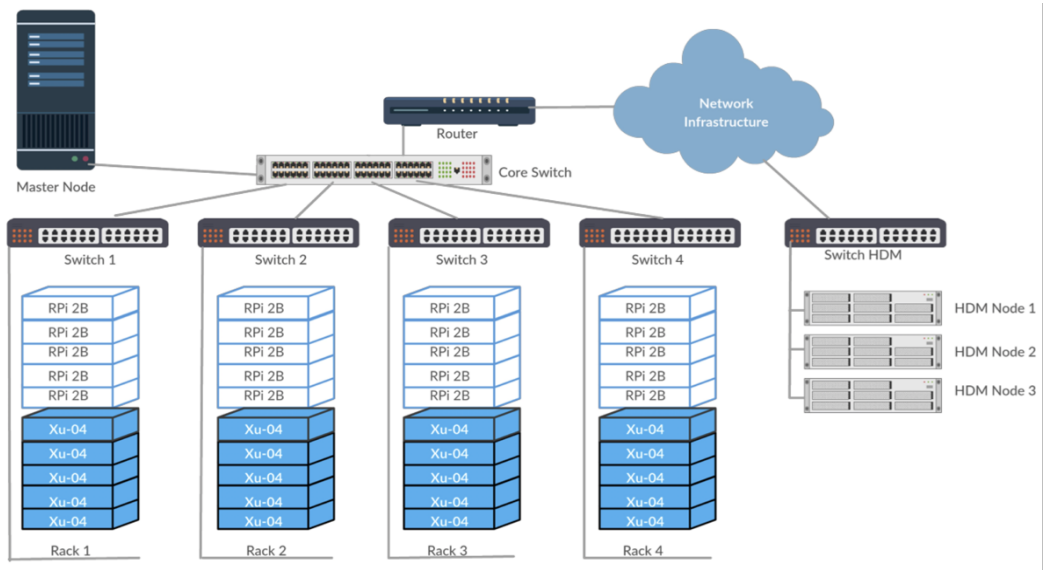

**Figure 1.** Network Topology diagram for RPi, Xu20 and HDM Clusters

2.6.2 is installed on the Image with Java JDK 7 for ARM platform. When ready, these SD cards are plugged into the RPi systems and mounted. The Master node is installed on a regular PC running an Ubuntu 14.4 virtual machine on Windows 10 as the host operating system.

For the Xu20 Cluster, we built another image based on Ubuntu MATE 15.10. Ubuntu MATE is an open source derivate of the Ubuntu Linux distribution with MATE desktop. HardKernel provides Ubuntu MATE 15.10 pre-installed on the Toshiba eMMCv5.0 memory module which is preconfigured for Odroid Xu-4 single board computers at a price of US$43. The eMMCv5.0 is capable of reading and write speeds of 140MB/s and 40MB/s respectively. Apache Hadoop 2.6.2 along with Java JDK 7 for ARM platform was installed on the image. These modules were inserted into eMMC socket on the Odroid Xu-4 boards and connected to the network. Similar to the RPi cluster, the Hadoop master node was installed on a regular PC running Ubuntu 14.4 VM. The final cluster HDM is composed of four PCs all connected in the same network topology as of the other clusters. A Virtual machine in the VMware workstation was built to run Hadoop 2.6.2 with Java JDK 7 for 64-bit architecture. One of the VMs serves as the master node and runs Hadoop namenode only. The rest of the VM run the data nodes of the cluster.

**Table 2.** Comparison of various features of the clusters

|  | RPi Cluster | Xu20 Cluster | HDM Cluster |
|---|---|---|---|
| Master Node | Intel i7 at 3.00 GHZ 64Bit Win 10 | Intel i7 at 3.00 GHZ 64Bit Win 10 | Intel i7 at 3.00 GHZ 64Bit Win 10 |
| Number of Data Nodes | 20 | 20 | 4 |
| Slave Node device | Raspberry Pi Model 2 B | HardKernel Odroid Xu-4 | Intel i7 at 3.00 GHZ 64Bit Win 10 |
| CPU Clock Speed | 1.0 GHz | 2.0 GHz | 3.0 GHz |
| OS | Raspbian OS | Ubuntu MATE 15 OS | Ubuntu 14.4 LTE |
| Storage (GB) | 16 GB | 32 GB | 40 GB |
| Storage Medium | Class 10 SD Card | eMMC 5.0 module | Kingston Solid State Disk (SSD) |
| RAM | 856 MB (available) | 1024 MB (available) | 3 GB (available) |
| Virtual Machine | Only Master Node runs OS in VM | Only Master Node runs OS in VM | All nodes on VM |

## 4. Performance Evaluation of DM-Clusters

In this section, we present a performance evaluation study of DM-clusters in terms of energy consumption, processing speed, storage read/write and networking.

*4.1. Energy consumption Approximation*

Resource over-provisioning and energy non-proportional behavior of today's servers [3,33,34] are two of the most important reasons for high energy consumption of data centers.

The Energy consumption for the DM-Clusters was measured using the Wattsup Pro .net power meters. These meters provide consumption in terms of watts for 24 hours a day and log these values in local memory for accessibility. To estimate the approximate power consumption over a year, we measured the power consumption in two modes, Idle-mode and Stress-mode for each DM-Cluster. In idle mode, the clusters were deployed without any application/task running for a period of 24 hours. In Stress-mode, the clusters ran a host of computation intensive applications for a period of 24 hours. Observing the logs, the upper-bound wattage usage within a period of 23 hours was taken as power consumption in the idle-mode as well as the stress-mode. Table 3 shows the power consumption for DM-Clusters in Idle and Stress Modes.

**Table 3.** Power consumption of clusters in idle and stress modes with Power cost per year

|  | Idle mode | | Stress mode | |
|---|---|---|---|---|
|  | Power consumption (E) | Power Cost in USD | Power consumption (E) | Power Cost in USD |
| RPi Cluster (20 nodes) | 34.1 W | $14.94 | 46.4 W | $20.33 |
| XU20 Cluster (20 nodes) | 56.2 W | $24.63 | 78.7 W | $34.49 |
| HDM Cluster (4 nodes) | 108.4 W | $47.51 | 197.7 W | $86.66 |

The cost of energy for the cluster is a function of power consumption per year and the cost of energy per kilo-Watts hour. An approximation of energy consumption cost per year ($C_y$) can be given by the Equation (1) where E is the specific power consumption for an event for 24 hours a day and 365.25 days per year. The approximate cost for all the clusters computed based on values given in Table 3; whereas the cost per kilowatt-hour (P) is assumed to be 0.05 US$.

$$C_y = E \times 24 \frac{hours}{day} \times 365.25 \frac{days}{year} \times \frac{P}{kWh} \quad (1)$$

The Bolzano Experiment [16] report raspberry Pi cluster built using Raspberry Pi Model B (first generation) where each node is consuming 3 Watts in stress mode. In RPi Cluster the Raspberry Pi Model 2B consumes slightly less power with 2.4W in stress mode. We observe that this slight difference in power consumption is due to the improved design of the second generation Raspberry Pi. The Cardiff Cloud testbed reported in [35] compare two Intel Xeon based servers deployed in the data center with each server consisting of 2 Xeon e5462 CPU (4 cores per processor), 32 GB of main memory and 1 SATA disk of 2 TB of storage each. The researchers in this study used similar equipment to measure power consumption as presented in this study. Their work reports that each server on average consumes 115W and 268W power in idle and stress modes respectively. The power consumption for the RPi Cluster with 20 nodes is 5 times better compared to a typical server in a cluster.

In a scenario where the RPi Cluster runs an application in stress mode (i.e. 46.4 W) for the whole year, the cost for power usage is approximately $20.33. For Xu20 and HDM Clusters, the yearly cost would be $34.49 and $86.66 respectively. Given these values, we can hypothesize that running a SBC based cluster would be cheaper and good for a greener computing environment in terms of energy consumption.

### 4.2. CPU Performance

The benchmark suite Sysbench was used to measure the CPU performance. Sysbench provides benchmarking capabilities for Linux and supports testing CPU, memory, File I/O, mutex performance in clusters. We execute the Sysbench benchmark, testing each number up to value 10,000 if it is a prime number for n number of threads. Since each computer has a quad-core processor we run the sysbench CPU test for 1, 2, 4, 8 and 16 threads. We measure the performance of this benchmark test for Raspberry Pi Model 2B, Odroid Xu-4 as well as Intel i7 4th Generation Computers used in the three DM-clusters. Table 4 shows the average CPU execution time for nodes with n threads.

Table 4. CPU Execution Time (seconds) for individual nodes with n threads

|  | CPU Cores | Clock rate Ghz | CPU Execution Time with n threads | | | | |
|---|---|---|---|---|---|---|---|
|  |  |  | 1 | 2 | 4 | 8 | 16 |
| Raspberry Pi 2B | 4 | 1.0 | 448.2 | 225.1 | 113.8 | 113.7 | 113.7 |
| Odroid Xu-4 | 8 | 2.0 | 83.3 | 41.68 | 25.33 | 17.66 | 18.02 |
| Intel i7 4th Gen | 4 | 3.0 | 8.51 | 4.272 | 2.22 | 2.27 | 2.23 |

All the tested devices had four cores, the CPU execution times scale well with the increased number of threads. Sysbench test runs with n=2 and n=4 threads significantly improve the execution times performance for all processors by 50%. With n=8 and n=16 threads, the test results yield almost similar execution times with little improvement in performance. We observe, the execution times for Odroid Xu-4 are 10 times better as compared to Raspberry Pi Model 2B. The increased number of threads does not provide gain in performance of Odroid Xu-4 over Raspberry Pi, furthermore, the execution time for Raspberry Pi is further extended with larger n. The HDM Cluster nodes run 4.42 times faster compared to Odroid Xu-4. These results clearly illustrate the handicap of SBC on-board processors when compared to a typical PC.

The Raspberry Pi Model 2B allows the user to overclock the CPU rate to 1200 MHz, in our experiments with the over-clocked CPU we did not observe significant improvement using the sysbench benchmark.

### 4.3. Storage Performance

Poor storage read/write performance can be a bottleneck in clusters. Compared to server machines, an SBC is handicapped in terms of availability of limited storage options. The small scale of the SBCs of Odroid Xu-4, as well as Raspberry Pi Model 2B, provide few options for external storage. The Raspberry Pi's were equipped with 16GB SanDisk Class 10 SD Cards, whereas the XU-4 devices were equipped with 32GB eMMC memory cards. Both of these memory cards were loaded with bootable Linux distributions. For comparison purposes, we used 128GB SanDisk Solid State Disks on the HDM Cluster machines and used flexible IO (FIO).FIO allows benchmarking of sequential read and write as well as random read and write with various block sizes. NAND memory is typically organized in pages and groups with sizes 4, 8 or 16 Kilobytes. Although it is possible for a controller to overwrite pages, the data cannot be overwritten without having to erase it first. The typical erase block on SD Cards is typically 64 or 128KB. As a result of these design features, the random read and write performance of SD Cards depends on the erase block, segment size, the number of segments and controller cache for address translations.

Table 5 shows the comparison of buffered and non-buffered random read and write from all the three devices with block size 4KB. FIO was used to measure the random read and write throughput with 8 threads each working with a file of size 512MB with a total 4GB of data. These parameters were set specifically to avoid buffering and caching in RAM issues which are managed by the underlying operating systems that can distort the results; i.e. the data size (4GB) selected is larger than the onboard RAM available on these devices. As can be seen from the Table 5, the read throughput (buffered) of Odroid with eMMC memory is at least twice as fast as the Class 10 SDCard on the Raspberry Pi whereas the non-buffered read is more than three times better. Similarly, for buffered write operations, Odroid Xu-4 with eMMC module throughput is more than twice better when compared to the Class 10 SDCard in Raspberry Pi. The buffered read throughput for SSD storage is at least 10 times better compared to eMMC module in Odroid Xu-4 computers whereas the buffered write throughput of SSD storage is 15 times better. These experimental observations clearly imply the benefit of using SSDs with higher throughput when compared to Class 10 SD cards as well as eMMC v5.0 memory modules.

Table 5. Read and Write Throughput (KB/sec) for individual devices in the clusters using FIO.

|  | Read Throughput (KB/second) | | Write Throughput (KB/second) | |
|---|---|---|---|---|
|  | Buffered | Non-Buffered | Buffered | Non-Buffered |
| Raspberry Pi 2B with 16 GB Class 10 SanDisk SDCard | 7,135 | 4,518 | 2,701 | 2,537 |
| Odroid Xu-4 with 32GB eMMCv5.0 Module | 14,318 | 13,577 | 6,421 | 5,118 |
| Intel i7 4th Gen with 120GB SanDisk Solid State Disk | 164,521 | 93,608 | 96,987 | 62,039 |

### 4.4. Network Performance

The network performance was measured using the popular Linux based command line tool iperf v3.13 with the NetPIPE benchmark version 3.7.2. After 30 test-runs, iperf states the network throughput to be 82-88 Mbits per second for the RPi and XU20 Clusters. NetPIPE, on the other hand, provides more details considering performance aspects for network latency, throughput etc. over a range of messages with various payload size in bytes. For this study, we executed the benchmark within the clusters for various payload sizes over the TCP end-to-end protocol. The NPtcp, NetPIPE benchmark using TCP protocol, involves running transmitter and receiver on two nodes in the cluster. In our experimentation, we executed the receiver on the cluster namenode with 1000 KB as maximum transmission buffer size for a period of 240 milliseconds. The transmitter was executed on the individual SBCs one by one. As can be seen from figure 2, the network latency for all clusters with small payload is almost similar. As the payload increases, we observe a slight increase in network latency between the three clusters. On the other hand, we observe a spike in throughput at message size 1000 bytes, this indicates that the smaller a message is, the more is the transfer time dominated by the communication layer overhead.

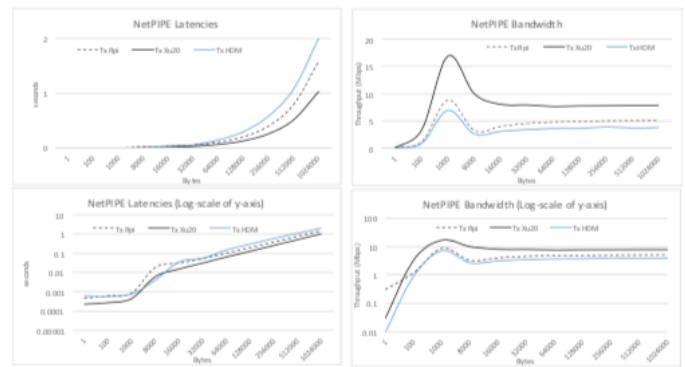

Figure 2. NetPIPE benchmark results for all clusters considering latencies and bandwidth with data size in terms of bytes on the x-axis.

Contrasting the performance of Xu-4 and RPi SBCs we note the visible difference in throughput between the two, this is due to the poor overall Ethernet performance of the Raspberry Pi probably caused by design. On the Raspberry Pi, 10/100Mbps Ethernet controller is a component of the LAN9512 controller which contains the USB 2.0 hub as well as the 10/100 Mbit Ethernet controller. On the other hand, the Odroid Xu-4 is equipped with an onboard Gigabit Ethernet controller which is part of the RTL8153 controller. The coupling of faster Ethernet port with high-speed USB 3.0 provides better network performance. Figure 2 also shows the comparison of throughput on the Xu20 cluster which is 1.52 times better than the RPi cluster.

## 5. Performance of Hadoop Benchmark Tests on Clusters

Apache Hadoop is an open source framework that provides distributed processing of large amounts of data in a data center [34,36,37]. On all three clusters, Hadoop version 2.6.2 was installed due to the availability of YARN daemon [34] which improves the performance of the map-reduce jobs in the cluster. To optimize the performance of these Clusters, yarn-site.xml and Mapred-site.xml were configured with 852 MB of resource size allocation. The primary reason for this is the limitation in the RPi Model 2B which has 1 GB of onboard RAM out of which 852MB is available; the rest is used by the Operating System. The default container size on the Hadoop Distributed File System (HDFS) is 128 MB. Each SBC node was assigned a static IPv4 address based on the configuration and all slave nodes were registered in the Master node. YARN and HDFS containers and interfaces could be monitored using the web interface provided by Hadoop [38]. Table 6 provide details of important configuration properties for the Hadoop environment. It must be noted that maximum memory allocation per container is 852MB; this is set on purpose so that the performance of all clusters could be measured and contrasted. Additionally, the replication factor for HDFS was set to 2.

Table 6. Properties in Hadoop configuration files

| Properties (mapred-site.xml) | Value |
| --- | --- |
| yarn.app.mapreduce.am.resource.mb | 852 |
| mapreduce.map.cpu.vcores | 1 |
| mapreduce.reduce.cpu.vcores | 1 |
| mapreduce.map.memory.mb | 852 |
| mapreduce.reduce.memory.mb | 852 |
| mapreduce.input.fileinputformat.split.minsize | 8 MB |
| Properties (YARN-site.xml) | Value |
| yarn.nodemanager.resource.memory-mb | 1024 |
| yarn.nodemanager.resource.cpu-vcores | 1 |
| yarn.scheduler.minimum-allocation-mb | 256 |
| yarn.scheduler.maximum-allocation-mb | 852 |
| yarn.scheduler.minimum-allocation-vcores | 1 |
| yarn.scheduler.maximum-allocation-vcores | 1 |
| yarn.nodemanager.vmem-pmem-ratio | 2 |
| Properties (hdfs-site.xml) | Value |
| dfs.replication | 2 |

These clusters were tested extensively for performance using Hadoop benchmarks such as DFSIO, TeraGen, TeraSort as well as Quasi-Random Pi generation and word count applications.

*5.1. The Pi Computation Benchmark*

Hadoop provides its own benchmarks for performance evaluation over multiple nodes. We execute the compute pi program on the clusters. The precision value m is provided at the command prompt with values ranging from $1\times10^3$ to $1\times10^6$ increased at an interval of $1\times10^1$. Each of these is run against a number of map tasks set at 10 and 100. We study the impact of the value of m versus the number of map tasks assigned and compute the difference in time consumption (execution time) for completion of these tasks. Each experiment is repeated at least 10 times for significance of statistical analysis. In this experimentation, the Pi computation benchmark's goal is to observe the CPU bound workload of all the three clusters. Figure 3 (a) and (b) shows the box-whisker plot with upper and lower quartiles for each sample set with 10 and 100 map tasks respectively. With 10 maps, the average execution time for RPi Cluster with $10^6$ number of samples is 100.8 seconds, whereas for XU20 and HDM Cluster the average execution time is 38.2 and 25.1 seconds respectively. As the number of maps increases to 100, we observe significant degradation in performance of RPi cluster with average execution time at 483.7 seconds for $10^6$ number of samples. Comparatively, the execution times for Xu20 and HDM Clusters are 50.1 and 21.8 seconds respectively. This clearly shows the significant difference in the computation performance between the RPi Cluster and the Xu20 Cluster. Figure 3 (c) shows the ratio of performance degradation of RPi and XU20 clusters compared to HDM cluster for Pi program CPU execution times with 10 and 100 maps.

*5.2. Wordcount Benchmark*

The Wordcount program contained in the Hadoop distribution is a popular micro-benchmark widely used in the community [39]. The Wordcount program is representative of a large subset of real-world MapReduce jobs extracting a small amount of interesting data from a large dataset. The Wordcount program reads text files and counts how often words occur within the selected text files. Each mapper takes a line from a text file as input and breaks it into words. It then emits a key/value pair of the word and a count value. Each reducer sums the count values for each word and emits a single key/value pair containing the word itself and the sum that word appears in the input files.

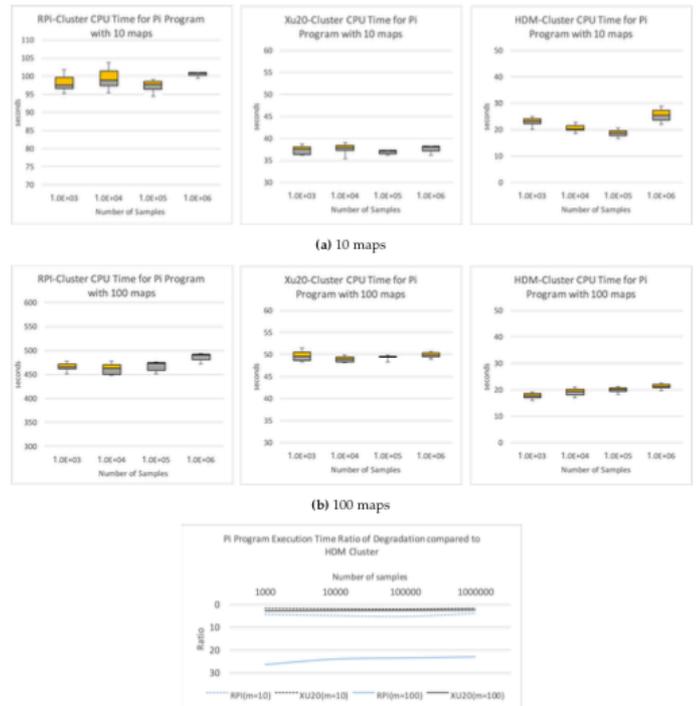

Figure 3. CPU Execution Time versus number of m samples for Computation of Pi benchmark in all clusters.

In our experimentation, we generated three large files of sizes 3, 30 and 300 Megabytes, respectively. Each experiment was run on the clusters separately at least 10 times for statistical accuracy. Figure 4 (a) shows the performance of CPU execution time, for the Wordcount benchmark for all clusters against input files sizes 3, 30 and 300 MB, in seconds on a logarithmic scale. The RPi Cluster performs 4 times worse compared to the Xu20 cluster and 12.5 times worse compared to HDM Cluster. The effect of the slower clock speed of the processor in the RPi nodes is clearly evident with smaller input file sizes of 3 MB. The average execution times of RPi and XU20 should be comparable since Wordcount generates only one mapper for each run resulting in a single container read by the mapper; however, the slower storage throughput with SD Cards adds to the overall latency. With input file size 30MB, Wordcount generates 4 mappers reading four containers from different nodes in the cluster, increasing the degree of parallelization thus reducing the overall CPU execution time.

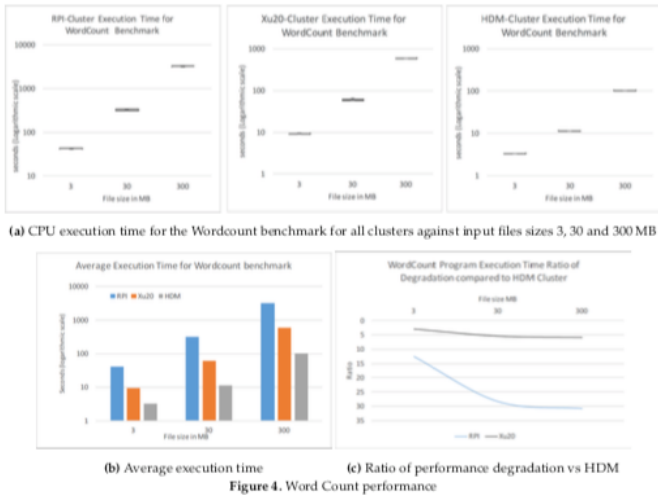

Figure 4. Word Count performance
(a) CPU execution time for the Wordcount benchmark for all clusters against input files sizes 3, 30 and 300 MB
(b) Average execution time
(c) Ratio of performance degradation vs HDM

Finally, with 300MB as input file size, we observe execution time performance correlating with smaller data sets, although the increased numbers of mappers should have improved the overall execution time. This is due to the fact that Wordcount generated 36 mappers for the job since there are only 19 nodes available (1 reserved for reducing job) in the Xu20 and RPi Clusters, the rest of the mappers would queue for the completion of previous mapper jobs resulting in increased overhead and reduced performance. Figure 4 (b) shows the average CPU execution times for all three clusters with different input file sizes. Furthermore, we observe that the Wordcount program executing on Xu20 is 2.8 times slower compared to HDM cluster for file size 3MB. For larger file sizes Xu20 is over 5 times slower compared to HDM Cluster. RPi Cluster, on the other hand, performs worse from 12 to 30 times slower compared to the HDM Cluster.

*5.3. The TestDFSIO Benchmark*

TestDFSIO [40] is an HDFS benchmark included in all major Hadoop distributions. TestDFSIO is designed to stress test the storage I/O (read and write) capabilities of a Hadoop cluster. ITestDFSIO creates n mappers for n number of files to be created and read subsequently in parallel. The reduce tasks collect and summarize the performance values. The test provides I/O performance information by writing a set of files of a fixed size to HDFS and subsequently reading these files while measuring Average I/O rate (MB/second), throughput (MB/second) and execution time (seconds) for the job. Since TestDFSIO requires files to be written first before they can be read, we run experiments to write 10 files of varying sizes for each experiment on all clusters. Each experiment is executed 5 times to obtain accurate results.

We consider the execution time of the TestDFSIO write benchmark with 10 files of sizes 1, 5, 10 and 20 GB. We observe that the execution time for the RPi cluster increases by 50% for file sizes 5GB and larger. The results for RPi cluster is correlating with Xu20 cluster albeit the execution time is less than half for the later. Comparatively, for the HDM Cluster, the execution time increases as the file size increases. As HDM Cluster consists of only 4 nodes, the replication factor increases the read/write operations to the nodes in the cluster causing increased network activity, therefore, increasing network latency issues.

We observe that the throughput improves as we increase the file size from 1GB to 10GB for RPi and Xu20 clusters. For larger file size (20GB) the throughput for Xu20 improves further whereas it degrades for RPi cluster. We also note that for the HDM Cluster the throughput decreases as the file size increases beyond 5GB. On average, the throughput for HDM cluster is better compared to the other clusters. For larger file sizes, the HDM cluster creates a number of blocks per HDFS node compared to RPi and Xu20 clusters; this is due to the less number of nodes in the HDM cluster causing increased write activity resulting in decreased write throughput. On the other hand, for the Xu20 cluster, the DFSIO write throughput is at least 2.2 times better compared to the RPi Cluster, as can be seen in figure 5 (a).

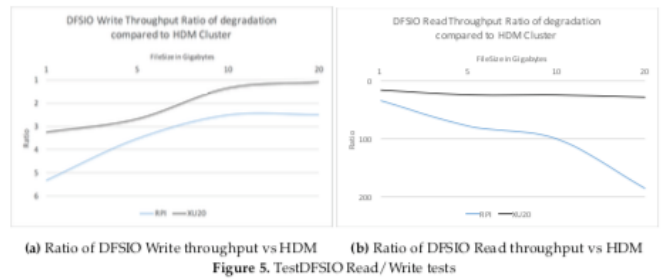

(a) Ratio of DFSIO Write throughput vs HDM
(b) Ratio of DFSIO Read throughput vs HDM
Figure 5. TestDFSIO Read / Write tests

We use the TestDFSIO Read test after completion of the written test. The read test reads the output files written to the HDFS by the previous test and observes execution time, throughput and average I/O rate. We measure the results using 10 files of sizes 1, 5, 10 and 20 GB and run experiments 5 times each. We note that the read performance of the HDM Cluster in terms of execution time is 15 and 33 times better against Xu20 Cluster and RPi cluster respectively. The performance degrades as the file size increases for all clusters. In contrast to these results, we observe that the read throughput for RPi cluster decreases by 68%, 70% and 15% with file sizes 5, 10 and 20GB. On the other hand, the read-through performance improves for the HDM as well as Xu20 clusters. It can be noted that RPi cluster's read performance degrades for large file sizes (20GB) whereas it is stagnant for Xu20 Cluster when compared to HDM Cluster. Figure 5 (b) shows the ratio of DFSIO Read throughput of RPi as well as Xu20 Clusters against the HDM Cluster. Table 7 shows the CPU Execution times, throughput and Average IO for TestDFSIO Read and Write Benchmark on clusters for various file sizes.

We observe, the average IO Rate, and the throughput for DFSIO write, increases for Xu20 Cluster and the RPi cluster, whereas it decreases for the HDM cluster.

*5.4. TeraSort Benchmark*

The Hadoop TeraSort benchmark suite sorts data as fast as possible to benchmark the performance of the MapReduce framework[41,42]. TeraSort combines testing the HDFS and MapReduce layers of a Hadoop cluster and consists of three MapReduce programs, TeraGen, TeraSort, and TeraValidate. TeraGen is typically used to generate large amounts of data blocks. This is achieved by running multiple concurrent map tasks. In our experimentation, we use TeraGen to generate large datasets to be sorted using a number of map tasks writing 100-byte rows of data to the HDFS. TeraGen divides the desired number of rows by the desired number of tasks and assigns ranges of rows to each map. Consequently, TeraGen is a write-intensive I/O benchmark.

Table 7. CPU Execution times, throughput and Average IO for TestDFSIO Read and Write Benchmark on clusters

| | File size (GB) | Avg. CPU Execution Times (seconds) | | | Average Throughput (mb/s) | | | Average IO rate (mb/s) | | |
|---|---|---|---|---|---|---|---|---|---|---|
| | | RPi Cluster | Xu20 Cluster | HDM Cluster | RPi Cluster | Xu20 Cluster | HDM Cluster | RPi Cluster | Xu20 Cluster | HDM Cluster |
| Read | 1 | 102.00 | 38.87 | 20.98 | 0.7 | 1.15 | 3.76 | 0.71 | 1.27 | 3.94 |
| | 5 | 98.64 | 37.84 | 20.35 | 1.35 | 1.78 | 4.79 | 1.44 | 2.16 | 5.91 |
| | 10 | 98.53 | 38.5 | 23.01 | 1.15 | 2.82 | 3.81 | 1.63 | 2.88 | 4.75 |
| | 20 | 107.32 | 38.96 | 30.80 | 1.34 | 3.09 | 3.36 | 1.50 | 2.90 | 3.99 |
| Write | 1 | 93.94 | 39.19 | 18.24 | 4.88 | 10.59 | 162.1 | 5.31 | 44.23 | 231.55 |
| | 5 | 94.35 | 37.4 | 17.37 | 7.16 | 23.63 | 555.65 | 8.87 | 62.6 | 626.63 |
| | 10 | 93.79 | 39.05 | 17.92 | 6.95 | 29.34 | 693.83 | 8.71 | 73.61 | 728.33 |
| | 20 | 96.19 | 40.64 | 21.09 | 4.87 | 33.1 | 904.19 | 6.75 | 77.42 | 930.27 |

In our experimentation, we run TeraGen, TeraSort and TeraValidate on all three clusters for various runs with data size in the range of 100MB, 200MB, 400MB, 800MB and 1.6GB respectively. We observe the job execution time for each run for comparison and analyze the performance on each cluster. The experiments were run 15 times for each data-size on each cluster. Table 8 shows the completion time CPU Execution Time for TeraGen in RPi, Xu20 and HDM Clusters for varying data payloads. Performance in terms of job completion time is correlating in all three clusters when payloads are increased, however, the completion time for HDM cluster is much faster in comparison. Similarly, when we contrast the TeraGen performance for Xu20 cluster against the RPi cluster, Xu20 clearly performs better. Since TeraGen is I/O intensive, the write speeds of the memories/storage in corresponding nodes in the clusters play a major role in degrading the overall job completion time. Table 8 also shows the job completion time for all clusters using TeraSort. The input data for TeraSort was previously generated by TeraGen in 100MB, 200MB, 400MB, 800MB and 1.6GB datasets respectively. The input data was previously written to the HDFS. For all experiments, we use the same number of map and reduce tasks on each cluster. The TeraSort benchmark is CPU bound during the map phase, i.e. reading input files and sorting tasks are carried out whereas it is I/O bound during the reduce phase, i.e. writing output files in the HDFS. We observed that 33-39% of job completion time occurred in map phase while 53% or more time spent in reduce tasks overall for the majority of TeraSort jobs run on all clusters. The HDM Cluster's TeraSort job completion time was observed to be 10 times faster when compared to RPi Cluster for all dataset payloads. The Xu20 Cluster's job execution time is at least 2.84 times better compared to RPi Cluster for all payloads.

Table 8. Average CPU Execution times for TeraSort and TeraGen Benchmarks on the clusters

| Data-set size (MB) | CPU Time for TeraGen (seconds) | | | CPU Time for TeraSort (seconds) | | |
|---|---|---|---|---|---|---|
| | RPi Cluster | Xu20 Cluster | HDM Cluster | RPi Cluster | Xu20 Cluster | HDM Cluster |
| 100 | 53.76 | 17.64 | 3.24 | 232.80 | 78.90 | 19.62 |
| 200 | 99.60 | 32.29 | 6.37 | 467.81 | 152.43 | 35.44 |
| 400 | 188.10 | 61.87 | 13.83 | 933.20 | 308.32 | 72.50 |
| 800 | 360.68 | 116.93 | 16.51 | 1872.43 | 607.42 | 138.88 |
| 1600 | 700.61 | 232.65 | 38.40 | 36614.1 | 1287.9 | 278.89 |

## 6. Power consumption and Temperature

We study the power consumption on all clusters using the TeraGen and TeraSort benchmark due to their intensive CPU and IO bound operations. As mentioned earlier in our cluster setup, we use a virtual machine to run the master node of the cluster which executes the namenode as well as the YARN ResourceManager Hadoop applications. The slave nodes execute the datanodes as well as the YARN NodeManager tasks. In order to avoid the influence of the namenode which is run as a virtual machine on a PC, we attach the power measurement equipment to the clusters slave nodes only and collect power consumption data. The WattsUp Pro .net meter is capable of recording power consumption in terms of watts, each reading is collected every second and is logged in the meter's onboard memory. The meters are initialized 10 seconds before each TeraGen and TeraSort job is initiated and stops reading 10 seconds after the job is completed. In addition to power consumption readings, we also periodically measured (every minute) the CPU temperature (Celsius) for both RPi's as well as Odroid Xu-4 boards in the cluster.

Figure 6 (a) shows the comparison of power consumption and CPU Temperature for both clusters for the TeraGen using 400MB datasets. It can be noted, the power consumption for Xu20 cluster peaks at 71.9 watts whereas RPi cluster consumes at most 46.3 watts. The temperature on RPi SBC mostly stays within the range 29-32C, on the other hand, Odroid Xu-4 SBC's are equipped with a cooling fan. At 45C, the fan turns on due to the built-in hardware settings yielding to increased power consumption on the Odroid Xu-4. Since TeraGen is IO bound job, initially mappers start executing and writing to the HDFS, as the progress continues some of the mappers complete the tasks assigned. Consequently, we a reduction in the overall power consumption of the cluster, this effect can be clearly observed in figure 6 (a) with 400MB Data size and figure 6 (b) with 800MB Data Size for both clusters. Figure 6 (c) and (d) shows the power consumption for both cluster when TeraSort is used. We observe that TeraSort requires more time for completion. Initially, mappers read through the input files generated by TeraGen and stored in HDFS, as the TeraSort shuffle process for keys and values initiates, we observe increased power consumption which continues until the mappers as well as the majority of reduce jobs complete. As the mappers continue to complete the tasks, the incoming results start processing in the reduce jobs. Before the completion of all map functions, the reduce functions initiate sorting and summarizing process requiring CPU as well as IO resources towards completion of the tasks.

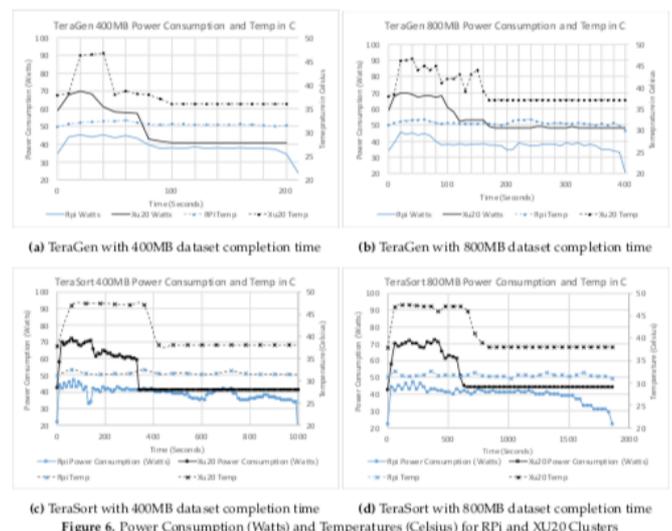

(a) TeraGen with 400MB dataset completion time
(b) TeraGen with 800MB dataset completion time
(c) TeraSort with 400MB dataset completion time
(d) TeraSort with 800MB dataset completion time
Figure 6. Power Consumption (Watts) and Temperatures (Celsius) for RPi and XU20 Clusters

We plot the percentage of the map and reduce completion against the power consumption for RPi Cluster with 400 and 800MB data size in figures 7 (a) and (b) respectively and for Xu20 Cluster in figures 7 (c) and (d). As can be seen, the percentage of maps and reduces completed correlates with the power consumption. In particular, when the map and reduce complete, the power consumption decreases, therefore, highlighting underutilized nodes in the clusters. Both TeraGen and TeraSort exhibit different power consumption. TeraSort on both clusters has a relatively long phase of higher power consumption from initialization of map jobs until about 80% of map jobs completion indicating high CPU utilization.

Afterwards, the power consumption decreases slightly fluctuating while both map and reduce jobs are executing in parallel. Finally, the power consumption steadies with minor tails and peaks in the plot towards reduce jobs completion. We observe that the trends for power consumption relevant to task completion are similar for larger data sizes used in this study and is consistent to observations in [34].

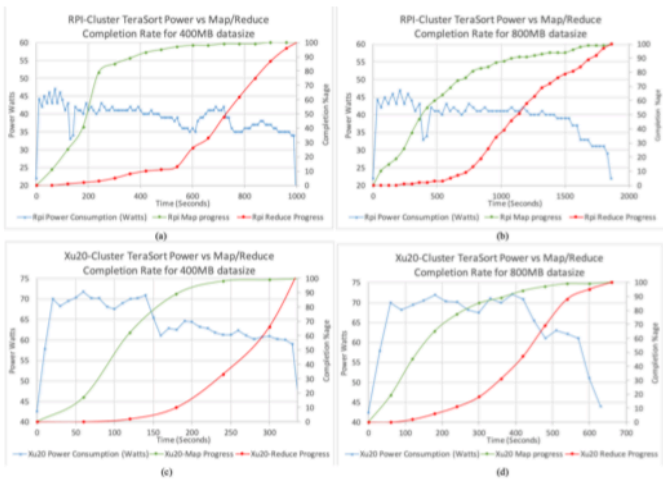

Figure 7. TeraSort Power consumption for (a) RPi Cluster with 400MB datasize (b) RPi Cluster with 800MB datasize (c) Xu20 Cluster with 400MB datasize. (d) Xu20 Cluster with 800MB datasize. In all plots, the secondary axis shows completion of map and reduce jobs in percentage.

## 7. Discussion

In this paper, we conducted an extensive study with varying parameters on the Hadoop cluster deployed using ARM-based single board computers. An overview of popular ARM-based SBCs Raspberry Pi, as well as HardKernel Odroid Xu-4 SBCs, was presented. The work also detailed the capabilities of these devices and tested them using popular benchmarking approaches. Details on requirements, design, and architecture of clusters built using these SBCs were provided. Two SBC clusters based on RPi and Xu-4 devices were constructed in addition to a PC based cluster running in the Virtual environment. Popular Hadoop benchmark programs such as Wordcount, TestDFSIO, and TeraSort were tested on these clusters and their performance results from the benchmarks were presented. This section presents a discussion of our findings and main lessons learned.

Deployment of Clusters: Using low-cost SBCs is an amicable way of deploying a Hadoop cluster at a very affordable cost. The low-cost factor would encourage students to build their own clusters, to learn about installation, configuration and operation of a cloud computing test-beds. The cluster also provides a platform for developers to build applications, test and deploy in public/private cloud environments. The small size of the SBCs allows installation of up to 32 nodes in a single module for a 1U rack mounting form factor. Further to this, these small clusters can be packaged for mobility and can be deployed in various emergency and disaster recovery scenarios.

Hadoop configuration optimization: Section 4 presents a comparison of CPU execution times using sysbench for both SBCs considered in this paper. Xu-4 devices in Xu20 Cluster perform better due to higher clock speeds and larger onboard RAM. Using sysbench we observed that increasing the number of cores in the CPU intensive benchmark, the execution time decreased. In Hadoop deployment configuration, we noticed that increasing the number of cores resulted in RPi cluster to be irresponsive for heavier workloads. On the other hand, Xu-4 boards performed well with an increased number of cores (up to 4). A possible explanation for this behavior is the Hadoop deployment setting where each core is assigned 852MB of memory, additional cores running Hadoop tasks would have to request virtual memory from the slower SD Cards resulting in poor performance leading to responsiveness. Although RPi devices are equipped with quad-core processors, due to the poor performing SD-Cards, it is inadvisable to use multiple-cores for Hadoop deployment.

In Hadoop deployment, not all of the available RAM onboard SBCs was utilized since we only allow one container to execute in YARN Daemon. The size of the container was set to 852MB which is the maximum available onboard memory in a Raspberry Pi node. This was intentionally done in order to study the performance correlation with the similar amount of resources in both kinds of SBCs. In further experimentation, we notice that Xu-4 devices are capable of handling up to four containers in each core at a time, resulting in better performance. We will further investigate the performance of all cores on the SBCs using Hadoop deployment of larger replication factors and a large number of YARN containers executing per node. On the HDM cluster running Hadoop environment in a Virtual machine, we note that higher replication factors resulted in a large number of errors due to replication overheads resulting in Hadoop stuck in an unrecoverable state. The SD Cards are slow and the storage provided per node in the cluster is distributed over the network degrading the overall performance of the cluster. Raspberry Pi with slower network port at speeds 10/100 Mbps also poses a considerable degradation in network performance. On the other hand, Xu20 cluster performed well comparatively with faster eMMC memory modules on board the Xu-4 devices. The SSD storage used in the HDM Cluster on the PCs provide the best performance in terms of storage IO although the network configuration of this cluster was a hindrance. We will consider using Network Attached Storage (NAS) attached to the master node where every rack would have a dedicated volume managed by Logical Volume Manager (LVM) that would be shared by all SBCs in the clusters.

Table 9. Average CPU Execution times for TeraSort and TeraGen Benchmarks on the clusters

| Bench-mark | Test Parameters | CPU Execution Time (seconds) | | | Energy Consumption (Watts) per job | | | Cost (Dollars) per job | | |
|---|---|---|---|---|---|---|---|---|---|---|
| | | RPi Cluster | Xu20 Cluster | HDM Cluster | RPi Cluster | Xu20 Cluster | HDM Cluster | RPi Cluster | Xu20 Cluster | HDM Cluster |
| | | | | | 10 maps | | | | | |
| Pi Program | $10^3$ samples | 98.47 | 37.37 | 22.86 | 1.27 | 0.817 | 1.26 | 0.063500 | 0.000040 | 0.000062 |
| | $10^4$ samples | 99.13 | 37.69 | 20.50 | 1.28 | 0.824 | 1.13 | 0.063900 | 0.000040 | 0.000056 |
| | $10^5$ samples | 97.90 | 36.97 | 18.92 | 1.26 | 0.808 | 1.04 | 0.063100 | 0.000040 | 0.000052 |
| | $10^6$ samples | 100.63 | 37.87 | 25.35 | 1.30 | 0.828 | 1.39 | 0.064800 | 0.000040 | 0.000069 |
| | | | | | 100 maps | | | | | |
| | $10^3$ samples | 465.68 | 49.62 | 17.84 | 6.002 | 1.08 | 0.980 | 0.300000 | 0.000054 | 0.000049 |
| | $10^4$ samples | 461.40 | 49.70 | 19.35 | 5.95 | 1.09 | 1.06 | 0.297000 | 0.000054 | 0.000053 |
| | $10^5$ samples | 470.26 | 49.43 | 20.12 | 6.06 | 1.08 | 1.10 | 0.303000 | 0.000054 | 0.000055 |
| | $10^6$ samples | 486.48 | 49.89 | 21.24 | 6.27 | 1.09 | 1.17 | 0.314000 | 0.000054 | 0.000058 |
| Word count | Filesize = 3MB | 41.12 | 9.25 | 3.29 | 0.53 | 0.202 | 0.181 | 0.000026 | 0.000010 | 0.000009 |
| | 30MB | 318.75 | 59.75 | 11.22 | 4.11 | 1.31 | 0.616 | 0.000205 | 0.000065 | 0.000032 |
| | 300MB | 3131.60 | 588.45 | 101.38 | 0.404 | 0.129 | 5.57 | 0.002000 | 0.000643 | 0.000278 |
| DFS IO write | Dataset = 1 | 102.00 | 38.87 | 20.98 | 1.31 | 0.85 | 1.15 | 0.000065 | 0.000042 | 0.000057 |
| | 5 | 98.64 | 37.84 | 20.35 | 1.27 | 0.827 | 1.12 | 0.000063 | 0.000041 | 0.000056 |
| | 10 | 98.53 | 38.50 | 23.01 | 1.27 | 0.842 | 1.26 | 0.000063 | 0.000042 | 0.000063 |
| | 20 | 107.32 | 38.96 | 30.80 | 1.38 | 0.852 | 1.69 | 0.000069 | 0.000042 | 0.000084 |
| DFS IO read | Dataset = 1 | 93.94 | 39.19 | 18.24 | 1.21 | 0.857 | 1.00 | 0.000060 | 0.000042 | 0.000051 |
| | 5 | 94.35 | 37.40 | 17.37 | 1.22 | 0.818 | 0.954 | 0.000061 | 0.000041 | 0.000047 |
| | 10 | 93.79 | 39.05 | 17.92 | 1.21 | 0.854 | 0.984 | 0.000060 | 0.000042 | 0.000049 |
| | 20 | 96.19 | 40.64 | 21.09 | 1.24 | 0.888 | 1.16 | 0.000062 | 0.000044 | 0.000057 |
| TeraGen | Dataset = 100 | 53.76 | 17.64 | 3.24 | 0.693 | 0.386 | 0.178 | 0.346000 | 0.000019 | 0.000008 |
| | 200 | 99.60 | 32.29 | 6.37 | 1.28 | 0.706 | 0.35 | 0.000064 | 0.000035 | 0.000017 |
| | 400 | 188.10 | 61.87 | 13.83 | 2.42 | 1.35 | 0.759 | 0.000121 | 0.000067 | 0.000038 |
| | 800 | 360.68 | 116.93 | 16.51 | 4.65 | 2.56 | 0.907 | 0.000232 | 0.000128 | 0.000045 |
| | 1600 | 700.61 | 232.65 | 38.40 | 9.03 | 5.09 | 2.11 | 0.000452 | 0.000254 | 0.000010 |
| | Dataset = 100 | 232.80 | 78.90 | 19.62 | 3.00 | 1.72 | 1.08 | 0.000150 | 0.000086 | 0.000053 |

Power efficiency: A motivation for this study was to analyze the power consumption of SBC based clusters. Due to their small form factor, SBC devices are inherently energy efficient, it is worth investigating if a cluster comprising of SBCs as nodes provides a better performance ratio in terms of power consumption and dollar cost. Although we did not measure the FLOPs per watt efficiency of either of our clusters, we notice wide inconsistencies in energy consumption results reported in the literature [4–8,38] for similar devices. This is due to the power measurement instruments varying results and inconsistencies in the design of power supplies. RPi, as well as Xu-4 devices, have no standard power supply, and micro USB based Power supply with unknown efficiency can be used. Since the total power consumed in the cluster is small, the efficiency of power supplies can make a big difference in overall power consumption. Nonetheless, WattsUp meters were effectively used to observe and analyze the power utilization for each task over the period of its execution in all experimentation. It is difficult to monitor and normalize the energy consumption for every test run over a period of time. It was observed that the MapReduce jobs, in particular, tend to consume more energy initially while map tasks are created and distributed across the cluster, while a reduction in power consumption is observed towards the end of the job. For the computation of power consumption, we assumed max power utilization (stress mode) for each job, during a test run in the clusters. Based on the power consumption of each cluster and the dollar cost of maintaining the clusters (given in table 3), a summary of average execution times, energy consumption and cost of running various benchmark tasks is presented in table 9.

Figure 8 (a) shows the energy consumption (in watts) for all Hadoop benchmarks with lowest workloads. Although the power consumption of RPi cluster is the lowest, the overall energy consumption by RPi cluster is the highest compared to Xu20 and HDM clusters due to the time inefficiency in job completion. In particular, with TeraSort benchmark which requires higher CPU and IO work rate, RPi and Xu20 clusters consume 2.7 and 1.6 times more energy compared to the HDM cluster for each TeraSort job. It is also worth noting that, apart from TeraGen and TeraSort, Xu20 Cluster proved to be, on average, 15-18% more energy efficient when compared to other clusters. This trend continues even for larger workloads as can be seen in figure 8 (b), Xu20 cluster is more energy efficient compared to RPi and HDM clusters for all Hadoop benchmarks with the exception of TeraGen and TeraSort. Results from these studies show that while SBC based clusters are energy efficient overall, the operation cost to performance ratio can vary based on the workload. For heavier workload application, such as big data applications, due to the inefficient performance, SBC based clusters may not be an appropriate choice.

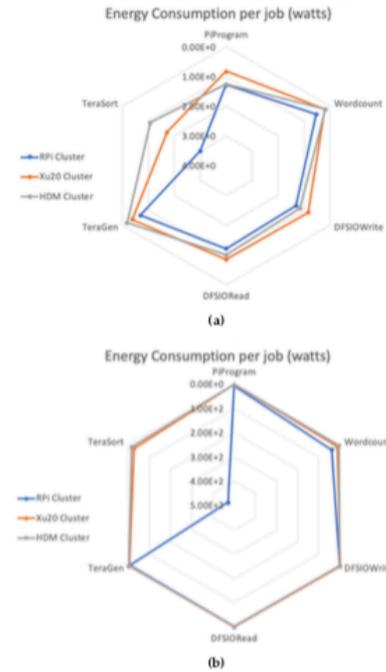

Figure 8. Energy consumption in clusters for all benchmarks with (a) lowest parameter settings, (b) highest parameter settings.

Cost of operating SBC based clusters: As mentioned in section 4A, the deployment cost of SBC based clusters is only a fraction of a traditional cluster composed of high-end servers. On the other hand, it is crucial to study the comparative cost of operating these clusters while considering the cluster's performance execution time as a factor. The dollar cost of execution of a task was computed using equation (1) where the cost is a function of task execution time and power consumption of the cluster. This approach has been used in literature [43–45]. Based on this, a detailed energy consumption in terms of watts and operation cost in terms of dollars per job is given in table 9. Figure 9 shows the ratio of operating cost of RPi and Xu20 clusters against the HDM cluster with smaller and larger workloads. For the benchmarks including Pi, DFSIORead and write; on average the RPi cluster is 2.63 and 4.45 times slower than the Xu20 and HDM clusters. The RPi cluster is almost always more expensive to operate compared to HDM cluster due to longer job completion times. The Xu20 cluster is less expensive for all benchmarks except TeraGen and TeraSort, where the cost could be as high as 100% compared to the HDM cluster. Based on these results, it can be concluded that while the cost of deployment of SBC based clusters is very low, the overall cost of operation can be expensive mainly due to the inefficient onboard SBC resources resulting in larger execution times for job completion effectively ensuing increased operation costs.

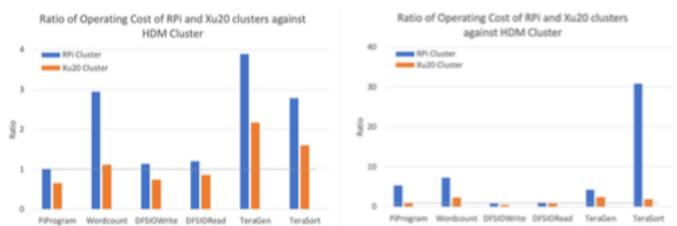

Figure 9. Ratio of operating cost (dollars) of RPi and Xu20 Clusters against HDM Cluster with (a) lowest parameter settings (b) highest parameter settings.

## 8. Conclusions and future work

This work investigates the role of SBC based clusters in energy efficient data centers in the context of big data applications. Hadoop was deployed on two low-cost low power ARM-based SBC clusters using Raspberry Pi and Odroid Xu-4 platforms.

We conduct a thorough experimental evaluation of the clusters comparing the performance parameters using popular benchmarks for CPU execution times, I/O read write, network I/O and power consumption. Further to this, we compare the clusters using Hadoop specific benchmarks including Pi computation, Wordcount, TestDFSIO, and TeraSort. An in-depth analysis of energy consumption of these clusters for various workloads is provided.

Results from these studies show that while SBC based clusters are energy efficient overall, the operation cost to performance ratio can vary based on the workload. For Smaller workloads the results shows that Xu20 cluster costs 32% and 152% less (in dollars) to operate compared to the HDM and RPi clusters. In terms of power efficiency, for smaller workloads, the Xu20 cluster outperforms the other clusters. For low-intensity workloads, the Xu20 cluster fares 37% better than the HDM cluster; however, the TeraGen and TeraSort heavy workloads yield higher energy consumption for Xu20 cluster with 2.41 and 1.84 times higher when compared to HDM cluster. The RPi cluster, for all performance benchmarks, yield poor results compared to the other clusters.

The cost of executing a large workload on a SBC-based cluster can be expensive mainly due to the limited on board resources on SBCs. These result in larger execution times for job completion effectively ensuing larger operation costs. It is, however, possible to tweak Hadoop configuration parameters on these clusters to improve the overall cost of operation. The low cost benefit of using SBC clusters is an attractive opportunity in green computing. These computers are increasingly becoming powerful and may help improve the energy efficiency in data centers. In future, we intend to study the application of SBC clusters on the edge of the cloud.

**Acknowledgement:** This work is supported by the Research and Innovation Center through grant number SSP-18-5-01. This work is also partially supported by the Robotics and Internet of Things Lab. The equipment used was provided by King Abdulaziz City for Science and Technology (KACST) through the Grant Number 157-37.